\documentstyle[amssymb,12pt]{article}

\newlength{\minitwocolumn}
\font\teneufm=eufm10
\font\seveneufm=eufm7
\font\fiveeufm=eufm5
\newfam\eufmfam
\textfont\eufmfam=\teneufm
\scriptfont\eufmfam=\seveneufm
\scriptscriptfont\eufmfam=\fiveeufm

\makeatletter
\@addtoreset{equation}{section}
\makeatother

\title{\bf The Baxter's $Q$-operator \\
for the $W$-algebra $W_N$\\
}
\begin{document}
\maketitle
\begin{center}
{
Takeo KOJIMA}
\\~\\
{\it
Department of Mathematics,
College of Science and Technology,
Nihon University,\\
Surugadai, Chiyoda-ku, Tokyo 101-0062, 
JAPAN
}
\end{center}
\begin{abstract}
The $q$-oscillator representation
for the Borel subalgebra of the affine symmetry
$U_q'(\widehat{sl_N})$ is presented.
By means of this $q$-oscillator representation,
we give the free field realizations of
the Baxter's $Q$-operator ${\bf Q}_j(\lambda),
\overline{\bf Q}_j(\lambda), (j=1,2,\cdots,N)$
for the $W$-algebra $W_N$.
We give the functional relations of the
$T$-$Q$ operators,
including the higher-rank generaliztion of the Baxter's
$T$-$Q$ relation.
\end{abstract}
~\\
~\\
\begin{center}
Key Words : CFT, 
$q$-oscillator, $Q$-operator, functional relation,
free field realization
\end{center}

\newpage

\section{Introduction}

The Baxter's $T$-$Q$-operator have various exceptional properties
and play an important role in many aspect of the theory of 
integrable systems.
Originally the $Q$-operator
was introduced by R.Baxter
\cite{Baxter1}, in terms of some special transfer matrix
of the 8-vertex model.
Over the last three decades,
this method of the $Q$-operator 
has been developed by many literatures.
We would like to refer some of these literatures,
written by R.Baxter \cite{Baxter2, Baxter3, Baxter4, Baxter5},
by L.Takhtadzhan and L.Faddeev \cite{TF},
by K.Fabricius and B.McCoy \cite{FM1,FM2,FM3},
by K.Fabricius \cite{Fab},
by V.Bazhanov and V.Mangazeev \cite{BM},
by B.Feigin, T.Kojima, J.Shiraishi and H.Watanabe \cite{FKSW2},
by T.Kojima and J.Shiraishi \cite{KS}.
However a full theory of the $Q$-operator for the 8-vertex
model is not yet developed. 
For the simpler models associated 
with the quantum group $U_q(g)$,
there have been many papers which extend,
generalize, and comment on the $T$-$Q$ relation.
We would like to refer 
some of these literatures, including Sklyanin's separation variable method,
written by E.Sklyanin 
\cite{Skl1, Skl2, Skl3}, 
by V.Kuzunetsov, V.Mangazeev and E.Sklyanin \cite{KMS}, 
by V.Pasquier and M.Gaudin \cite{PG},
by S.Derkachov \cite{Der1}
by S.Derkachov, G.Karakhanyan and A.Mansahov \cite{DKM1,DKM2}
by S.Derkachov, G.Karakhanyan and R.Kirschner, \cite{DKK}
by S.Derkachov and A.Mansahov \cite{DM},
by A.Belisty, S.Derkachov, G.Korchemesky and A.Manasahov \cite{BDKM},
by C.Korff \cite{Kor1,Kor2},
by A.Bytsko and J.Teschner \cite{BT},
by V.Bazhanov, S.Lukyanov and Al.Zamolodchikov \cite{BLZ1,BLZ2,BLZ3,BLZ4},
by M.Rossi and R.Weston \cite{RW},
by P.Dorey and R.Tateo \cite{DT},
by V.Bazhanov, A.Hibberd and S.Khoroshkin \cite{BHK},
by P.Kulish and Z.Zeitlin \cite{KZ},
by A.Antonov and B.Feigin \cite{AF},
by I.Krichever, O.Lipan, P.Wiegmann and A.Zabrodin \cite{KLWZ},
by V.Bazhanov and N.Reshetikhin \cite{BR},
by A.Kuniba, T.Nakanishi and J.Suzuki \cite{KNS},
by H.Boos, M.Jimbo, T.Miwa, F.Smirnov and Y.Takeyama \cite{BJMST1, BJMST2},
by A.Chervov and G.Falqui \cite{CF}.
Each paper added to our understanding of the  
great Baxter's original paper \cite{Baxter1}.
Especially for example the $T$-$Q$-operators 
acting on the Fock space
of the Virasoro algebra $Vir$ were introduced by
V.Bazhanov, S.Lukyanov and Al.Zamolodchikov \cite{BLZ1,
BLZ2, BLZ3}.
They derived various functional relations
of the $T$-$Q$ operators and gave
the asymptotic behavior of the eigen-value of the
$T$-$Q$ operators.
P.Dorey and R.Tateo \cite{DT} 
revealed the hidden connection between 
the vacuum expectation value of the $Q$-operator and
the spectral determinant for Schr\"odinger equation.
V.Bazhanov, A.Hibberd and S.Khoroshkin \cite{BHK}
achieved the $W_3$-algebraic generalization of
\cite{BLZ1,BLZ2,BLZ3,BLZ4,DT}.
In this paper 
we study the higher-rank $W_N$-generalization of
\cite{BHK}.
We study the $T$-$Q$-operators acting on the Fock space
of the $W$-algebra $W_N$.
We give the free field realization of the $Q$-operator 
and functional relations of
the $T$-$Q$-operators for
the $W$-algebra $W_N$,
including the higher-rank generalization of
the Baxter's $T$-$Q$ relation,
\begin{eqnarray}
&&
{\bf Q}_i(tq^N)+\sum_{s=1}^{N-1}(-1)^{s}
{\bf T}_{\Lambda_1+\cdots+
\Lambda_s}(tq^{-1}){\bf Q}_i(tq^{N-2s})+(-1)^{N}{\bf Q}_i(tq^{-N})=0,
\nonumber
\\
&&
\overline{\bf Q}_i(tq^{-N})+\sum_{s=1}^{N-1}(-1)^{s}
\overline{\bf T}_{\Lambda_1+\cdots+\Lambda_s}(tq)
\overline{\bf Q}_i(tq^{-N+2s})+(-1)^{N}
\overline{\bf Q}_i(tq^{N})=0,\nonumber
\end{eqnarray}
where $i=1,2,\cdots,N$.
The organization of this paper is as following.
In section 2, we give basic definitions, including
$q$-oscillator representation of the Borel subalgebra
of the affine symmetry $U_q'(\widehat{sl_N})$,
which play an essential role
in construction of the $Q$-operator.
In section 3, we give the definition of
the $T$-operator and the $Q$-operator.
In section 4 we give conjecturous funtional relations
between the $T$-opeartor and the $Q$-operator,
including Baxter's $T$-$Q$ relation.
In appendix, we give supporting arguments on
conjecturous formulae stated in section 4.

\section{Basic Definition}

In this section we give
the different realizations of the Borel subalgebra
of the affine quantum algebra $U_q'(\widehat{sl_N})$,
which will play an important role in
construction of the Baxter's
$T$-$Q$ operator.
Let us fix the integer $N \geqq 3$.
Let us fix a complex number
$1<r<N$.
In this paper,
upon this setting, we 
construct the Baxter's $T$-$Q$ operators
on the space of
the $W$-algebra $W_N$
with the central charge 
$-\infty<C_{CFT}<-2$, where
\begin{eqnarray}
C_{CFT}=(N-1)\left(1-\frac{N(N+1)}{r (r-1)}\right).\nonumber
\end{eqnarray}
Becuse $C_{CFT} \to -\infty$ represents the classical limit,
we call $-\infty<C_{CFT}<-2$ 
``quasi-classical domain''.
By anlytic continuation, it is possible 
to extend our theory to
the CFT with central charge $C_{CFT}<1$.
We would like to note that the unitary minimal CFT is
described by the central charge
$C_{CFT}=
(N-1)\left(1-\frac{N(N+1)}{r (r-1)}\right)$
for $N,r \in {\mathbb Z}$,
$(N \geqq 2, r\geqq N+2)$ \cite{FL}.
We set parameters $r^*=r-1$
and $q=e^{2\pi i \frac{r^*}{r}}$.
In what follows we use
the $q$-integer $[n]_q=\frac{q^n-q^{-n}}{q-q^{-1}}$.

\subsection{The $q$-oscillator representation}

Let $\{\epsilon_j\}$ be an orthonormal basis of ${\mathbb R}^N$,
relative to the standard inner product
$(\epsilon_i|\epsilon_j)=\delta_{i,j}$.
Let us set $\bar{\epsilon}_j=\epsilon_j-\epsilon$
where $\epsilon=\frac{1}{N}\sum_{j=1}^N \epsilon_j$.
We have $(\bar{\epsilon}_i|\bar{\epsilon}_j)
=\delta_{i,j}-\frac{1}{N}$.
Let us set 
the simple roots $\alpha_j=\bar{\epsilon}_j-\bar{\epsilon}_{j+1},
(1\leqq j \leqq N-1)$ and $\alpha_N=-\sum_{j=1}^{N-1}\alpha_j$.
Let us set 
the fundamental weights
$\omega_j$ as the dual vector of $\alpha_j$,
i.e. $(\alpha_i|\omega_j)=\delta_{i,j}$.
Explicitly we have $\omega_j=\bar{\epsilon_1}+\cdots+
\bar{\epsilon_j}$.
Let us set 
the weight lattice $P=\oplus_{j=1}^N {\mathbb Z}
\bar{\epsilon}_j$.
We consider the quantum affine algebra $U_q'(\widehat{sl_N})$,
which is generated by $e_1,\cdots,e_{N}$,
$f_1, \cdots, f_{N}$, and $h_1,\cdots, h_N$,
with the defining relations,
\begin{eqnarray}
&&~[h_i,h_j]=0,
~[h_i,e_j]=(\alpha_i|\alpha_j)e_j,
~[h_i,f_j]=-(\alpha_i|\alpha_j)f_j,
~[e_i,f_j]=\delta_{i,j}\frac{q^{h_i}-q^{-h_i}}
{q-q^{-1}},\nonumber\\
&&e_i^2 e_j-[2]_qe_i e_j e_i+e_j e_i^2=0,
~~
f_i^2 f_j-[2]_qf_i f_j f_i+f_j f_i^2=0,~~{\rm for}~~(\alpha_i|\alpha_j)=-1.\nonumber
\end{eqnarray}
Here $((\alpha_j|\alpha_k))_{1 \leqq j,k \leqq N}$
is the Cartan matrix of type $\widehat{sl_N}$.
Let us introduce the Borel subalgebra of $U_q'(\widehat{sl_N})$.
The Borel subalgebra $U_q'(\widehat{\bf b}^+)$
is generated by $e_1,\cdots,e_{N}$, 
$h_1,\cdots, h_{N}$,
and
$U_q'(\widehat{\bf b}^-)$
by $f_1,\cdots,f_{N}$, 
$h_1,\cdots,h_{N}$.
In this paper we consider the level $c=0$ case,
with the central element
$c=h_1+\cdots+h_{N}$.
Let us introduce the $q$-oscillator representation 
$o_t$ of
the Borel subalgebra $U_q'(\widehat{\bf b}^+)$.
The $q$-oscillator algebra $Osc_j$, $(1\leqq j \leqq N-1)$,
is generated by elements ${\cal E}_j, {\cal E}_j^*, {\cal H}_j$,
with the defining relations,
\begin{eqnarray}
~[{\cal H}_j,{\cal E}_j]={\cal E}_j,
~[{\cal H}_j,{\cal E}_j^*]=-{\cal E}_j^*,~
q{\cal E}_j{\cal E}_j^*-q^{-1}{\cal E}_j^*{\cal E}_j
=\frac{1}{q-q^{-1}}.
\end{eqnarray}
Let us set $Osc=Osc_1 \otimes_{{\mathbb C}} \cdots 
\otimes_{{\mathbb C}} Osc_{N-1}$.
We have $[{\cal E}_j,{\cal E}_k]=0$, 
$[{\cal E}_j^*,{\cal E}_k^*]=0$,
$[{\cal E}_j,{\cal E}_k^*]=0$,
$[{\cal H}_j,{\cal H}_k]=0$ for $j \neq k$.
Let us set the auxiliarry operator ${\cal H}_N=-{\cal H}_1
-{\cal H}_2-\cdots-{\cal H}_{N-1}$.
We define homomorphism 
$o_t: U_q'(\widehat{\bf b}^+) \to 
Osc$ by
\begin{eqnarray}
o_t(e_1)&=&t q^{\frac{1}{2}}(q-q^{-1})
q^{-{\cal H}_2}{\cal E}_1^*{\cal E}_2,
\nonumber\\
o_t(e_2)&=&q^{\frac{1}{2}}(q-q^{-1})
q^{-{\cal H}_3}{\cal E}_2^*{\cal E}_3,
\nonumber\\
\cdots \nonumber\\
o_t(e_{N-2})&=&q^{\frac{1}{2}}(q-q^{-1})q^{-{\cal H}_{N-1}}
{\cal E}_{N-2}^*{\cal E}_{N-1},\nonumber\\
o_t(e_{N-1})&=&{\cal E}_{N-1}^*,\nonumber\\
o_t(e_N)&=& q^{-{\cal H}_1-{\cal H}_N}{\cal E}_1,
\label{def:q-osc}
\end{eqnarray}
\begin{eqnarray}
o_t(h_1)=-{\cal H}_1+{\cal H}_2,~
o_t(h_2)=-{\cal H}_2+{\cal H}_3,\cdots,
o_t(h_N)=-{\cal H}_N+{\cal H}_1.\nonumber
\end{eqnarray}
This $q$-oscillator representation $o_t$ 
satisfies level zero condition
$o_t(h_1+h_2+\cdots+h_N)=0$.
This $q$-oscillator representation give a higher-rank generalization
of those in \cite{BHK}.
By means of the Dynkin-diagram automorphism
$\tau, \sigma$, we construct 
a family of the $q$-oscillator representation
$o_{t,j}, \bar{o}_{t,j}$.
Let us set the Dynkin-diagram automorphism 
$\tau$ of the affine algebra 
$U_q'(\widehat{sl_N})$.
\begin{eqnarray}
\tau(e_1)=e_2, \cdots, \tau(e_j)=e_{j+1},\cdots, \tau(e_N)=e_1,\nonumber\\
\tau(h_1)=h_2, \cdots, \tau(h_j)=h_{j+1},\cdots, \tau(h_N)=h_1,\nonumber\\
\tau(f_1)=f_2, \cdots, \tau(f_j)=f_{j+1},\cdots, \tau(f_N)=f_1.\nonumber
\end{eqnarray}
Let us set the Dynkin-diagram automorphism $\sigma$ of 
the finite simple algebra 
$U_q({sl_N})$, generated by
$e_2,\cdots,e_{N}, h_2,\cdots,h_{N}, f_2,\cdots, f_{N}$.
\begin{eqnarray}
\sigma(e_2)=e_{N},\cdots, \sigma(e_j)=e_{N+2-j},\cdots, \sigma(e_{N})=e_2, \nonumber\\
\sigma(h_2)=h_{N},\cdots, \sigma(h_j)=h_{N+2-j},\cdots, \sigma(h_{N})=h_2, \nonumber\\
\sigma(f_2)=f_{N},\cdots, \sigma(f_j)=f_{N+2-j},\cdots, \sigma(f_{N})=f_2, \nonumber
\end{eqnarray}
and $\sigma$ is esxtended to
the affine vertex as
$\sigma(e_1)=e_1,
\sigma(h_1)=h_1,
\sigma(f_1)=f_1$.
We have the action of $\tau^{j}\cdot \sigma \cdot 
\tau^{-1}$,
\begin{eqnarray}
\tau^{j}\cdot \sigma \cdot \tau^{-1}(e_{i})
=e_{j-1-i},\nonumber\\
\tau^{j}\cdot \sigma \cdot \tau^{-1}(h_{i})=
h_{j-1+i},\nonumber\\
\tau^{j}\cdot \sigma \cdot \tau^{-1}
(f_{i})=f_{j-1-i},\nonumber
\end{eqnarray}
with $s, j \in {\mathbb Z}$.
We set homomorphism $o_{t,j},\bar{o}_{t,j}: 
U_q'(\widehat{\bf b}^+) \to Osc$,
$(1\leqq j \leqq N)$, 
\begin{eqnarray}
o_{t,j}=o_t \cdot \tau^{-j},~~
\bar{o}_{t,j}=o_{(-1)^N t} \cdot \tau^{j} 
\cdot \sigma \cdot \tau^{-1},
\end{eqnarray}
These $q$-oscillator representations 
$o_{t,j},\bar{o}_{t,j}$
will play an important role in construction
of the Baxter's $Q$-operator.

\subsection{Evaluation highest weight representation}

Let us consider
the quantum simple algebra $U_q(gl_N)$,
which is generated by
$E_{\alpha_1},\cdots,E_{\alpha_{N-1}}$, 
$H_1,\cdots, H_{N}$,
and $F_{\alpha_1},\cdots, F_{\alpha_{N-1}}$,
with the defining relations,
\begin{eqnarray}
&&~[H_i,H_j]=0,
~[H_i,E_{\alpha_j}]=(\delta_{i,j}-\delta_{i,j+1})E_{\alpha_j},~
~[H_i,F_{\alpha_j}]=(-\delta_{i,j}+\delta_{i,j+1})F_{\alpha_j},
\nonumber\\
&&~[E_{\alpha_i},F_{\alpha_j}]=\delta_{i,j}\frac{q^{H_i-H_{i+1}}-q^{-H_i+H_{i+1}}}
{q-q^{-1}},\nonumber\\
&&E_{\alpha_i}^2 E_{\alpha_j}-[2]_qE_{\alpha_i} E_{\alpha_j} E_{\alpha_i}
+E_{\alpha_j} E_{\alpha_i}^2=0,
~~
F_{\alpha_i}^2 F_{\alpha_j}-[2]_qF_{\alpha_i} F_{\alpha_j} F_{\alpha_i}
+F_{\alpha_j} F_{\alpha_i}^2=0.\nonumber
\end{eqnarray}
Let us set the root vectors,
\begin{eqnarray}
&&F_{\alpha_1+\alpha_2}=[F_{\alpha_2},F_{\alpha_1}]_{\sqrt{q}}
=\sqrt{q}F_{\alpha_2}F_{\alpha_1}-\frac{1}{\sqrt{q}}
F_{\alpha_1}F_{\alpha_2},\nonumber\\
&&\bar{F}_{\alpha_1+\alpha_2}=[F_{\alpha_2},F_{\alpha_1}]_{
\frac{1}{\sqrt{q}}}
=\frac{1}{\sqrt{q}}
F_{\alpha_2}F_{\alpha_1}
-\sqrt{q}F_{\alpha_1}F_{\alpha_2},\nonumber\\
&&F_{\alpha_1+\cdots+\alpha_{N-1}}=[F_{\alpha_{N-1}},[F_{\alpha_{N-2}},
\cdots,[F_{\alpha_2},F_{\alpha_1}]_{\sqrt{q}}\cdots ]_{\sqrt{q}}]_{\sqrt{q}},\nonumber\\
&&\overline{F}_{\alpha_1+\cdots+\alpha_{N-1}}
=[[
\cdots,[F_{\alpha_{N-1}},F_{\alpha_{N-2}}]_{\frac
{1}{\sqrt{q}}}\cdots ,F_{\alpha_2}]_{\frac{1}{\sqrt{q}}},
F_{\alpha_1}]_{\frac{1}{\sqrt{q}}}.\nonumber
\end{eqnarray}
Let us set the automorhism $\sigma$ by
\begin{eqnarray}
&&\sigma(E_{\alpha_1})=E_{\alpha_{N-1}},\cdots,
\sigma(E_{\alpha_j})=E_{\alpha_{N-j}},\cdots,
\sigma(E_{\alpha_{N-1}})=E_{\alpha_{1}},\nonumber\\
&&\sigma(H_1)=-H_{N},\cdots,
\sigma(H_j)=-H_{N-j+1},\cdots,
\sigma(H_N)=-H_1,\nonumber\\
&&\sigma(F_{\alpha_1})=F_{\alpha_{N-1}},\cdots,
\sigma(F_{\alpha_j})=F_{\alpha_{N-j}},\cdots,
\sigma(F_{\alpha_{N-1}})=F_{\alpha_{1}}.\nonumber
\end{eqnarray}
We have the evaluation representation
$ev_t$,
$\overline{ev}_t$: 
$U_q'(\widehat{sl_N})\to U_q(gl_N)$, given by
\begin{eqnarray}
&&
ev_t(e_2)=E_{\alpha_1},\cdots, 
ev_t(e_{j+1})=E_{\alpha_j},\cdots, 
ev_t(e_{N})=E_{\alpha_{N-1}},\nonumber\\
&&
ev_t(h_2)=H_1-H_2,\cdots,
ev_t(h_{j+1})=H_j-H_{j+1},\cdots,
ev_t(h_{N})=H_{N-1}-H_N,\nonumber\\
&&
ev_t(f_2)=F_{\alpha_1},\cdots, 
ev_t(f_{j+1})=F_{\alpha_j},\cdots, 
ev_t(f_{N})=F_{\alpha_{N-1}},\nonumber
\\
&&
ev_t(e_1)=t
F_{\alpha_1+\alpha_2+\cdots+\alpha_{N-1}}q^{H_1+H_{N}},~
ev_t(f_1)=t^{-1}
E_{\alpha_1+\alpha_2+\cdots+\alpha_{N-1}}
q^{-H_1-H_{N}},\nonumber\\
&&ev_t(h_1)=H_N-H_1.\nonumber
\end{eqnarray}
\begin{eqnarray}
&&
\overline{ev}_t(e_2)=E_{\alpha_1},\cdots, 
\overline{ev}_t(e_{j+1})=E_{\alpha_j},\cdots, 
\overline{ev}_t(e_{N})=E_{\alpha_{N-1}},\nonumber\\
&&
\overline{ev}_t(h_2)=H_1-H_2,\cdots,
\overline{ev}_t(h_{j+1})=H_j-H_{j+1},\cdots,
\overline{ev}_t(h_{N})=H_{N-1}-H_N,\nonumber\\
&&
\overline{ev}_t(f_2)=F_{\alpha_1},\cdots, 
\overline{ev}_t(f_{j+1})=F_{\alpha_j},\cdots, 
\overline{ev}_t(f_{N})=F_{\alpha_{N-1}},\nonumber\\
&&
\overline{ev}_t(e_1)=t
\overline{F}_{\alpha_1+\alpha_2+\cdots+\alpha_{N-1}}
q^{-H_1-H_{N}},~
\overline{ev}_t(f_1)=t^{-1}
\overline{E}_{\alpha_1+\alpha_2+\cdots+\alpha_{N-1}}
q^{H_1+H_{N}},\nonumber\\
&&
\overline{ev}_t(h_{1})=H_{N}-H_1.\nonumber
\end{eqnarray}
We have the conjugation $\overline{ev}_t=
\sigma \cdot ev_{(-)^N t} \cdot \sigma^{-1}$.
We set the irreducible highest representation of 
$U_q(gl_N)$
with the highest weight 
$\lambda=m_1 \Lambda_1+\cdots+m_{N}\Lambda_{N}$,
the highest weight vector $|\lambda \rangle$
of $U_q(sl_N)$.
\begin{eqnarray}
&&
\pi^{(\lambda)}(E_{\alpha_j})|\lambda \rangle=0,~~
\pi^{(\lambda)}(H_{j})|\lambda \rangle=m_j|\lambda \rangle,
~~(1\leqq j \leqq N).\nonumber
\end{eqnarray}
In what follows we consider the case
$m_j-m_{j+1} \in {\mathbb N}$, $(1\leqq j \leqq N-2)$.
In this case 
the representation $\pi^{(\lambda)}$ is finite dimension.
Let us set the evaluation highest weight representation
$\pi_t^{(\lambda)}$ of 
the affine symmetry $U_q'(\widehat{sl_N})$,
as
\begin{eqnarray}
\pi^{(\lambda)}_t=\pi^{(\lambda)}\cdot ev_t,~~~
\overline{\pi}^{(\lambda)}_t=\pi^{(\lambda)}\cdot 
\overline{ev}_t.\nonumber
\end{eqnarray}
These evaluation highest weight representation 
will play an important role in construction
of the $T$-operator ${\bf T}_\lambda(t), 
\overline{\bf T}_\lambda(t)$.

\subsection{Screening current}

Let us introduce bosons $B_m^i$,
$(m \in {\mathbb Z}_{\neq 0}; i=1,2,\cdots,N-1)$ by
\begin{eqnarray}
[B_m^i,B_n^j]=m \delta_{m+n,0}(\alpha_i|\alpha_j)\frac{r-1}{r},~~
(1\leqq i,j \leqq N-1).
\end{eqnarray}
Let us set $B_m^N=-\sum_{j=1}^{N-1}B_m^j$.
We have the commutation relation
$
[B_m^i,B_n^j]=m \delta_{m+n,0}(\alpha_i|\alpha_j)\frac{r-1}{r}$,
for $1\leqq i,j \leqq N$.
Let us set the zero-mode operators $P_\lambda$ and $Q_\lambda$,
$(\lambda \in P=\oplus_j {\mathbb Z}\bar{\epsilon}_j)$ by
\begin{eqnarray}
[P_\lambda, iQ_\mu]=(\lambda |\mu).
\end{eqnarray}
Let us set the Heisenberg algebra ${\cal B}$
generated by $B_m^1,\cdots,B_m^{N-1}$,
$P_\lambda, Q_\lambda$, 
$(\lambda \in P)$ 
and its completion $\widehat{\cal B}$.
Let us set the Fock space ${\cal F}_{l,k}$ by
\begin{eqnarray}
B_m^j|l,k\rangle&=&0,~~(m>0)\\
P_\alpha|l,k\rangle&=&
\left(
\alpha \left|
\sqrt{\frac{r}{r-1}}l-
\sqrt{\frac{r-1}{r}}k
\right.
\right)
|l,k\rangle,\\
|l,k\rangle&=&e^{i\sqrt{\frac{r}{r-1}}Q_l-
i\sqrt{\frac{r-1}{r}}Q_k}|0,0\rangle.
\end{eqnarray}
Let us set the screening currents of the $W$-algebra
$W_N$ by
\begin{eqnarray}
V_{\alpha_j}(u)&=&
\exp\left(i\sqrt{\frac{r^*}{r}}Q_{\alpha_j}\right)
\exp\left(\sqrt{\frac{r^*}{r}}P_{\alpha_j}i u\right)\nonumber\\
&\times&
\exp\left(\sum_{m>0}\frac{1}{m}B_{-m}^j e^{imu}\right)
\exp\left(-\sum_{m>0}\frac{1}{m}B_m^j e^{-imu}\right),~
(1\leqq j \leqq N).
\end{eqnarray}
Here we have added one operator $V_{\alpha_N}(u)$,
which looks like affinization of 
the classical $A_{N-1}$.
We can find the elliptic deformation of
$V_{\alpha_j}(u)$ for $j \neq N$
in \cite{FKSW2, KS}.
For ${\rm Re}(u_1)>{\rm Re}(u_2)$, we have
\begin{eqnarray}
V_{\alpha_j}(u_1)V_{\alpha_j}(u_2)&=&
:V_{\alpha_j}(u_1)V_{\alpha_j}(u_2):(e^{iu_1}-e^{iu_2})^{\frac{2r^*}{r}},~~
(1\leqq j \leqq N),
\nonumber
\\
V_{\alpha_j}(u_1)V_{\alpha_{j+1}}(u_2)&=&
:V_{\alpha_j}(u_1)V_{\alpha_{j+1}}(u_2):
(e^{iu_1}-e^{iu_2})^{-\frac{r^*}{r}},~~
(1\leqq j \leqq N),\nonumber
\\
V_{\alpha_{j+1}}(u_1)V_{\alpha_j}(u_2)&=&
:V_{\alpha_{j+1}}(u_1)V_{\alpha_{j}}(u_2):
(e^{iu_1}-e^{iu_2})^{-\frac{r^*}{r}},~~
(1\leqq j \leqq N).
\nonumber
\end{eqnarray}
By analytic continuation, we have
\begin{eqnarray}
V_{\alpha_i}(u_1)V_{\alpha_j}(u_2)=q^{(\alpha_i|\alpha_j)}
V_{\alpha_j}(u_2)V_{\alpha_i}(u_1),
~~(1 \leqq i,j \leqq N).
\end{eqnarray}
Let us set
\begin{eqnarray}
z_{j}=\exp\left(-2 \pi i \sqrt{\frac{r^*}{r}}
P_{\bar{\epsilon}_{j}}\right),~~(1 \leqq j \leqq N).
\end{eqnarray}
We have $z_1 z_2\cdots z_{N}=1$ and
\begin{eqnarray}
V_{\alpha_i}(u+2\pi)
=
z_{i}^{-1}z_{i+1}V_{\alpha_i}(u),~~
z_i V_{\alpha_j}(u)&=&q^{\delta_{i,j+1}-\delta_{i,j}}
V_{\alpha_j}(u) z_i.\nonumber
\end{eqnarray}
Let us set the nilpotent subalgebra
$U_q'(\widehat{\bf n}^-)$ generated by
$f_1,f_2,\cdots,f_N$.
We have homomorphism 
$sc : U_q'(\widehat{\bf n}^-) \to \widehat{\cal B}$
given by
\begin{eqnarray}
sc(f_j)=\frac{1}{q-q^{-1}}\int_0^{2\pi}
V_{\alpha_j}(u)du,~~(1\leqq j \leqq N).\nonumber
\end{eqnarray}

\section{Baxter's $Q$-operator}

In this section we define
the Baxter's $T$-$Q$ operator by 
means of the trace of the universal $R$,
and present conjecturous functional relations of
the $T$-$Q$ operator, which include
the higher-rank generalization of the Baxter's $T$-$Q$ relation.

\subsection{${\cal L}$-operator}

Let us set the universal $L$-operator ${\cal L}
\in \widehat{\cal B}\otimes U_q(\widehat{\bf n}^-)$ by
\begin{eqnarray}
{\cal L}=\exp\left(-\pi i \sqrt{\frac{r^*}{r}}
\sum_{j=1}^{N}
P_{\omega_j} \otimes h_j
\right){\cal P}\exp\left(
\int_0^{2\pi}K(u)du\right).
\end{eqnarray}
Here we have set 
\begin{eqnarray}
K(u)=\sum_{j=1}^{N}
V_{\alpha_j}(u)\otimes e_j.\nonumber
\end{eqnarray}
Here ${\cal P}\exp
\left(\int_0^{2\pi}K(u)du\right)$
represents
the path ordered exponential
\begin{eqnarray}
{\cal P}\exp
\left(\int_0^{2\pi}K(u)du\right)=
\sum_{n=0}^\infty 
\int \cdots \int_{2\pi \geqq u_1 \geqq u_2 \geqq
\cdots \geqq u_n \geqq 0}
K(u_1)K(u_2)\cdots K_n(u_n)du_1 du_2 \cdots du_n. 
\nonumber
\end{eqnarray}
The above integrals converge in ``quasi-classical domain''
$-\infty < C_{CFT} <-2$.
For the value of $C_{CFT}$ outside the quasi-classical domain,
the integrals should be understood as analytic continuation.
Let us set $U_q(\widehat{sl_N})$ the extension of
$U_q'(\widehat{sl_N})$ by the degree operator $d$.
Let us set $U_q(\widehat{\bf n}^\pm)$ the extension of
$U_q'(\widehat{\bf n}^\pm)$ by the degree operator $d$.
There exists the unique universal $R$-matrix 
${\cal R} \in U_q(\widehat{\bf n}^+)\otimes
U_q(\widehat{\bf n}^-)$ satisfying
the Yang-Baxter equation.
\begin{eqnarray}
{\cal R}_{12}{\cal R}_{13}{\cal R}_{23}=
{\cal R}_{23}{\cal R}_{13}{\cal R}_{12}.\nonumber
\end{eqnarray}
The universal-$R$'s Cartan elements ${\bf t}$ is factored as
\begin{eqnarray} 
{\cal R}=q^{\bf t} \overline{\cal R},
~~
{\bf t}=
\sum_{j=1}^{N-1}h_j \otimes h^j+c \otimes d+d \otimes c,
\nonumber
\end{eqnarray}
where $(h^i|h_j)=\delta_{i,j}$.
We call the element $\overline{\cal R}
\in U_q'(\widehat{\bf n}^+)\otimes
U_q'(\widehat{\bf n}^-)$ the reduced universal $R$-matrix.
The ${\cal L}$-operator is an image of the
reduced $R$-matrix \cite{BHK},
\begin{eqnarray}
{\cal L}=(sc \otimes id )(\overline{R}).\nonumber
\end{eqnarray}
The ${\cal L}$-operator will play an important role
in trace construction of the $T$-$Q$ operator.

\subsection{T-operator}

Let us set the $T$-operator ${\bf T}_\lambda(t)$ and
$\overline{\bf T}_\lambda(t)$ by
\begin{eqnarray}
&&{\bf T}_\lambda(t)={\rm Tr}_{
\pi_t^{(\lambda)}}
\left(
\exp\left(-\pi i \sqrt{\frac{r^*}{r}}
\sum_{j=1}^N
P_{\omega_j} \otimes h_j
\right){\cal L}
\right),\\
&&\overline{\bf T}_\lambda(t)={\rm Tr}_{
\overline{\pi}_t^{(\lambda)}}
\left(
\exp\left(-\pi i \sqrt{\frac{r^*}{r}}
\sum_{j=1}^N
P_{\omega_j} \otimes h_j
\right){\cal L}
\right).
\end{eqnarray}
Let us set an image of ${\cal L}$ as
${\bf L}_\lambda(t)=(id \otimes \pi_t^{(\lambda)})
\left(
{\cal L}
\right)$, and the $R$-matrix $R_{\lambda_1,\lambda_2}(t_1/t_2)=
\pi_{t_1}^{(\lambda_1)}\otimes \pi_{t_2}^{(\lambda_2)}({\cal R})
$.
We have so-called $RLL$ relation,
\begin{eqnarray}
{R}_{\lambda_1, \lambda_2}(t_1/t_2)
{\bf L}_{\lambda_1}(t_1)
{\bf L}_{\lambda_2}(t_2)=
{\bf L}_{\lambda_2}(t_2)
{\bf L}_{\lambda_1}(t_1)
{R}_{\lambda_1, \lambda_2}(t_1/t_2).\nonumber
\end{eqnarray}
Multiplying the $R$-matrix
${R}_{\lambda_1, \lambda_2}(t_1/t_2)^{-1}$ from the right,
and taking trace, we have the commutation relation,
\begin{eqnarray}
~[{\bf T}_{\lambda_1}(t_1), {\bf T}_{\lambda_2}(t_2)]=
~[\overline{\bf T}_{\lambda_1}(t_1), 
\overline{\bf T}_{\lambda_2}(t_2)]=
~[{\bf T}_{\lambda_1}(t_1), \overline{\bf T}_{\lambda_2}(t_2)]=0.
\nonumber
\end{eqnarray}
The coefficients of the Taylor expansion of
${\bf T}_\lambda(t)$ commute with each other.
Hence we have infinitly many commutative operators,
which give quantum deformation of the conservation laws
of the $N$-th KdV equation.

\subsection{Q-operator}

Let us set the Fock representation $\pi_j^\pm$:
$Osc_j \to W^\pm$ with $j=1,2,\cdots,N-1$,
\begin{eqnarray}
W^+=\oplus_{k \geqq 0}{\mathbb C}|k\rangle_+,
~~
W^-=\oplus_{k \geqq 0}{\mathbb C}|k\rangle_-.\nonumber
\end{eqnarray}
The action is given by
\begin{eqnarray}
\pi_j^+({\cal H}_j)|k\rangle_+=-k|k\rangle_+,~
\pi_j^+({\cal E}_j)|k\rangle_+=\frac{1-q^{-2k}}{(q-q^{-1})^2}
|k-1\rangle_+,~
\pi_j^+({\cal E}_j^*)|k\rangle=|k+1\rangle_+,\nonumber\\
\pi_j^-({\cal H}_j)|k\rangle_-=k|k\rangle_-,~
\pi_j^-({\cal E}_j)|k\rangle_-
=\frac{1-q^{2k}}{(q-q^{-1})^2}|k-1\rangle_-,~
\pi_j^-({\cal E}_j)|k\rangle_-=|k+1\rangle_-.\nonumber
\end{eqnarray}
Let $\pi_j$ and $\overline{\pi}_j$ be any representation
of the $q$-oscillator
$Osc=Osc_1 \otimes_{\mathbb C}\cdots \otimes_{\mathbb C}
Osc_{N-1}$ such that the partition
$Z_j(t), \overline{Z}_j(t)$ converge. 
\begin{eqnarray}
Z_j(t)&=&{\rm Tr}_{\pi_j o_{t,j}}
\left(
\exp\left(-2 \pi i \sqrt{\frac{r^*}{r}}
\sum_{j=1}^{N}
P_{\omega_j} \otimes h_j
\right)
\right),\nonumber\\
\overline{Z}_j(t)&=&
{\rm Tr}_{\overline{\pi}_j \overline{o}_{t,j}}
\left(
\exp\left(-2 \pi i \sqrt{\frac{r^*}{r}}
\sum_{j=1}^{N}
P_{\omega_j} \otimes h_j
\right)
\right).\nonumber
\end{eqnarray}
Let us set the operators ${\bf A}_j(t)$ and
$\overline{\bf A}_j(t)$ with $j=1,2,\cdots,N$
\begin{eqnarray}
&&{\bf A}_j(t)=\frac{1}{Z_j(t)}
{\rm Tr}_{ 
{\pi}_j {o}_{t,j}
}
\left(
\exp\left(-\pi i \sqrt{\frac{r^*}{r}}
\sum_{j=1}^{N}
P_{\omega_j} \otimes h_j
\right){\cal L}
\right),\label{def:A1}
\\
&&\overline{\bf A}_j(t)=\frac{1}{\overline{Z}_j(t)}
{\rm Tr}_{
\overline{\pi}_j \overline{o}_{t,j}}
\left(
\exp\left(-\pi i \sqrt{\frac{r^*}{r}}
\sum_{j=1}^{N}
P_{\omega_j} \otimes h_j
\right){\cal L}
\right).\label{def:A2}
\end{eqnarray}
Let us set the Baxter's $Q$-operator ${\bf Q}_j(t)$ and
$\overline{\bf Q}_j(t)$ with $j=1,2,\cdots,N,$
\begin{eqnarray}
{\bf Q}_j(t)=t^{-\frac{1}{2}\sqrt{\frac{r}{r^*}}
P_{\bar{\epsilon}_{j}}}
{\bf A}_j(t),~~
\overline{\bf Q}_j(t)=
t^{\frac{1}{2}\sqrt{\frac{r}{r^*}}
P_{\bar{\epsilon}_{j}}}
\overline{\bf A}_j(t).\label{def:Q}
\end{eqnarray}
We would like to
note convenient relation,
\begin{eqnarray}
\sum_{k=1}^{N}P_{\omega_k}\otimes o_{t,j}(h_k)&=&
\sum_{k=1}^{N-1}(P_{\bar{\epsilon}_{j}}
-P_{\bar{\epsilon}_{j+k}})
\otimes {\cal H}_k,\nonumber\\
\sum_{k=1}^{N}P_{\omega_k}\otimes \overline{o}_{t,j}(h_k)
&=&
\sum_{k=1}^{N-1}(P_{\bar{\epsilon}_{j-k}}
-P_{\bar{\epsilon}_{j}})
\otimes {\cal H}_k.\nonumber
\end{eqnarray}
Here we should understand the surfix number as modulus $N$,
i.e. $\bar{\epsilon}_{j+N}=\bar{\epsilon}_j$.

From the Yang-Baxter equation, we have
the commutation relations
\begin{eqnarray}
~[{\bf Q}_{j_1}(t_1),{\bf Q}_{j_2}(t_2)]
=[\overline{\bf Q}_{j_1}(t_1),\overline{\bf Q}_{j_2}(t_2)]
=[{\bf Q}_{j_1}(t_1),\overline{\bf Q}_{j_2}(t_2)]=0,\nonumber
\end{eqnarray}
and
\begin{eqnarray}
~[{\bf Q}_{j}(t_1),{\bf T}_{\lambda}(t_2)]
=[{\bf Q}_{j}(t_1),\overline{\bf T}_{\lambda}(t_2)]
=[\overline{\bf Q}_{j}(t_1),{\bf T}_{\lambda}(t_2)]
=[\overline{\bf Q}_{j}(t_1),\overline{\bf T}_{\lambda}(t_2)]=0.\nonumber
\end{eqnarray}
The operators ${\bf A}_j(t)$ can be written as  power series.
\begin{eqnarray}
{\bf A}_j(t)&=&1+\sum_{n=1}^\infty
\sum_{\sigma_1,\cdots,\sigma_{Nn} \in {\mathbb Z}_N}
a_{Nn}^{(j)}(\sigma_1,\cdots,\sigma_{Nn})\nonumber\\
&\times&
\int \cdots \int_{2\pi \geqq u_1 \geqq u_2 \geqq \cdots
\geqq u_{Nn} \geqq 0}
V_{\alpha_{\sigma_1}}(u_1)
\cdots V_{\alpha_{\sigma_{Nn}}}(u_{Nn})
du_1 \cdots du_{Nn}.\nonumber
\end{eqnarray}
Here we have set
\begin{eqnarray}
a_{Nn}^{(j)}(\sigma_1,\cdots,\sigma_{Nn})=
\frac{1}{Z_j(t)}{\rm Tr}_{
{\pi}_j {o}_{t,j}}
\left(
\exp\left(-2\pi i \sqrt{\frac{r^*}{r}}
\sum_{j=1}^N
P_{\omega_j} \otimes h_j
\right)e_{\sigma_1}e_{\sigma_2}\cdots e_{\sigma_{Nn}}
\right).\nonumber
\end{eqnarray}
The coefficients $a_{Nn}^{(j)}$ 
vanishes unless
$
n=|\{ j |\sigma_j=s \}|$
for $s \in {\mathbb Z}_N$, and behaves like
$a_{Nn}^{(j)} \sim O(t^n)$.
The coefficients $a_{Nn}^{(j)}$
are determined by the commutation relations of
the Borel subalgebra $U_q(\widehat{\bf n}^-)$ and
the cyclic property of the trace,
hence the specific choice of representation
$\pi_j, \overline{\pi}_j$ is not significant
as long as it converges.
In \cite{FKSW2, KS}
we have constructed the elliptic version of
the integral of the currents,
$$
\int \cdots \int_{2\pi \geqq u_1 \geqq u_2 \geqq \cdots
\geqq u_{Nn} \geqq 0}
V_{\alpha_{\sigma_1}}(u_1)
V_{\alpha_{\sigma_2}}(u_2)
\cdots V_{\alpha_{\sigma_{Nn}}}(u_{Nn})
du_1 du_2 \cdots du_{Nn}.
$$

\section{Functional relations}

In the previous section, we show that the
$T$-$Q$ operators commute with each other.
In this section we give
conjecturous functional relations
of the $T$-$Q$ operators,
which coincide with the previous work
\cite{BHK} upon $N=3$ specialization.
We have checked those functional 
relations up to the order $O(t^2)$
in appendix.
Some of similar formulae have been obtained in the context of
the solvable lattice models associated with $U_q(\widehat{sl_N})$
\cite{KLWZ,BR,KNS}.
At the end of this section
we summarize conclusion.


\subsection{Functional relations}

The $T$-operator is written by determinant of the $Q$-operators.
Let us set the Young diagram $\mu=(\mu_1,\mu_2,\cdots,\mu_N)$,
$(\mu_j\geqq \mu_{j+1}; \mu_j \in {\mathbb N})$.
Using the same character as the Young diagram $\mu$,
we represent the highest weight 
$\mu=\mu_1 \Lambda_1+\cdots+\mu_{N}\Lambda_{N}$.
We set 
$c_0=\prod_{1\leqq j<k \leqq N}
\left(
\sqrt{\frac{z_j}{z_k}}-
\sqrt{\frac{z_k}{z_j}}
\right)$.
We have the following determinant formulae of the $T$-operator,
\begin{eqnarray}
{\bf T}_\mu(t)&=&
\frac{1}{c_0}\left|
\begin{array}{cccc}
{\bf Q}_1(tq^{2\tilde{\mu}_1})&
{\bf Q}_1(tq^{2\tilde{\mu}_2})&\cdots&
{\bf Q}_1(tq^{2\tilde{\mu}_N})\\
{\bf Q}_2(tq^{2\tilde{\mu}_1})&
{\bf Q}_2(tq^{2\tilde{\mu}_2})&\cdots&
{\bf Q}_2(tq^{2\tilde{\mu}_N})\\
\cdots&\cdots&\cdots&\cdots
\\
{\bf Q}_N(tq^{2\tilde{\mu}_1})&
{\bf Q}_N(tq^{2\tilde{\mu}_2})&\cdots&
{\bf Q}_N(tq^{2\tilde{\mu}_N})
\end{array}
\right|,
\label{rel:det1}\\
\nonumber\\
\nonumber\\
\overline{\bf T}_\mu(t)&=&
\frac{1}{c_0}
\left|
\begin{array}{cccc}
\overline{\bf Q}_1(tq^{-2\tilde{\mu}_1})&
\overline{\bf Q}_1(tq^{-2\tilde{\mu}_2})&\cdots&
\overline{\bf Q}_1(tq^{-2\tilde{\mu}_N})\\
\overline{\bf Q}_2(tq^{-2\tilde{\mu}_1})&
\overline{\bf Q}_2(tq^{-2\tilde{\mu}_2})&\cdots&
\overline{\bf Q}_2(tq^{-2\tilde{\mu}_N})\\
\cdots&\cdots&\cdots&\cdots
\\
\overline{\bf Q}_N(tq^{-2\tilde{\mu}_1})&
\overline{\bf Q}_N(tq^{-2\tilde{\mu}_2})&\cdots&
\overline{\bf Q}_N(tq^{-2\tilde{\mu}_N})
\end{array}
\right|.\label{rel:det2}
\end{eqnarray}
Here we have used the auxiliarry parameters 
$2\tilde{\mu}_j=2\mu_j+N-2j+1, (1\leqq j \leqq N)$.
We have checked the above formulae
(\ref{rel:det1}) and (\ref{rel:det2})
for $\mu=\Lambda_1$ and $\mu=
\Lambda_1+\cdots+\Lambda_{N-1}$,
up to the order $O(t^2)$. See appendix.
As the special case $\mu_j=0, (1\leqq j \leqq N)$, 
we have the quantum Wronskian condition.
\begin{eqnarray}
c_0&=&
\left|
\begin{array}{cccc}
{\bf Q}_1(tq^{N-1})&{\bf Q}_1(tq^{N-3})&\cdots&{\bf Q}_1(tq^{-N+1})\\
{\bf Q}_2(tq^{N-1})&{\bf Q}_2(tq^{N-3})&\cdots&{\bf Q}_2(tq^{-N+1})\\
\cdots&\cdots&\cdots&\cdots
\\
{\bf Q}_N(tq^{N-1})&{\bf Q}_N(tq^{N-3})&\cdots&{\bf Q}_N(tq^{-N+1})
\end{array}
\right|,
\label{rel:det1'}
\\
\nonumber
\\
\nonumber\\
c_0&=&
\left|
\begin{array}{cccc}
\overline{\bf Q}_1(tq^{-N+1})&
\overline{\bf Q}_1(tq^{-N+3})&\cdots&\overline{\bf Q}_1(tq^{N-1})\\
\overline{\bf Q}_2(tq^{-N+1})&\overline{\bf Q}_2(tq^{-N+3})&\cdots&
\overline{\bf Q}_2(tq^{N-1})\\
\cdots&\cdots&\cdots&\cdots
\\
\overline{\bf Q}_N(tq^{-N+1})&\overline{\bf Q}_N(tq^{-N+3})&\cdots&
\overline{\bf Q}_N(tq^{N-1})
\end{array}
\right|.\label{rel:det2'}
\end{eqnarray}
We have checked 
the above formulae 
(\ref{rel:det1'}), (\ref{rel:det2'}), 
up to the order $O(t^2)$. See appendix.
Let us set 
$c_i=\prod_{1\leqq j<k \leqq N
\atop{j,k \neq i}}
\left(
\sqrt{\frac{z_j}{z_k}}-
\sqrt{\frac{z_k}{z_j}}
\right)$ for $1\leqq i \leqq N$.
The two kind of $Q$-operator, ${\bf Q}_j(t)$ and
$\overline{\bf Q}_j(t)$,
are functionally dependent.
The $Q$-operator ${\bf Q}_i(t)$ is written by the determinant of
the $Q$-operator $\overline{\bf Q}_j(t)$,
\begin{eqnarray}
c_i {\bf Q}_i(t)&=&
\left|
\begin{array}{cccc}
\overline{\bf Q}_1(tq^{N-2})&\overline{\bf Q}_1(tq^{N-4})&\cdots&
\overline{\bf Q}_1(tq^{-N+2})\\
\cdots&\cdots&\cdots&\cdots
\\
\overline{\bf Q}_{i-1}(tq^{N-2})&
\overline{\bf Q}_{i-1}(tq^{N-4})&\cdots&\overline{\bf Q}_{i-1}(tq^{-N+2})\\
\overline{\bf Q}_{i+1}(tq^{N-2})&\overline{\bf Q}_{i+1}(tq^{N-4})&\cdots&
\overline{\bf Q}_{i+1}(tq^{-N+2})\\
\cdots&\cdots&\cdots&\cdots
\\
\overline{\bf Q}_N(tq^{N-2})&
\overline{\bf Q}_N(tq^{N-4})&\cdots&\overline{\bf Q}_N(tq^{-N+2})
\end{array}
\right|,
\label{rel:det3}
\\
\nonumber
\\
\nonumber
\\
c_i 
\overline{\bf Q}_i(t)&=&
\left|
\begin{array}{cccc}
{\bf Q}_1(tq^{-N+2})&{\bf Q}_1(tq^{-N+4})&\cdots&{\bf Q}_1(tq^{N-2})\\
\cdots&\cdots&\cdots&\cdots
\\
{\bf Q}_{i-1}(tq^{-N+2})&{\bf Q}_{i-1}(tq^{-N+4})&\cdots&{\bf Q}_{i-1}(tq^{N-2})\\
{\bf Q}_{i+1}(tq^{-N+2})&{\bf Q}_{i+1}(tq^{-N+4})&\cdots&{\bf Q}_{i+1}(tq^{N-2})\\
\cdots&\cdots&\cdots&\cdots
\\
{\bf Q}_N(tq^{-N+2})&{\bf Q}_N(tq^{-N+4})&\cdots&{\bf Q}_N(tq^{N-2})
\end{array}
\right|,
\label{rel:det4}
\end{eqnarray}
with $i=1,2,\cdots,N$.
We have checked the determinat formulae (\ref{rel:det3}) 
and (\ref{rel:det3}) up to the order $O(t^2)$.
See appendix.
We derive the following
(\ref{rel:BaxTQ1}), (\ref{rel:BaxTQ2}), 
(\ref{rel:qua1}), (\ref{rel:qua2}),
(\ref{rel:qua3}), (\ref{rel:qua4}),
 and (\ref{rel:JT})
from the above formulae
(\ref{rel:det1}), (\ref{rel:det2}), 
(\ref{rel:det3}) and (\ref{rel:det4}).
We have the higher-rank generalization of
the Baxter's $T$-$Q$ relation (\ref{rel:BaxTQ1})
and (\ref{rel:BaxTQ2}), as the consequence of
(\ref{rel:det1}) and (\ref{rel:det2}),
\begin{eqnarray}
&&
{\bf Q}_i(tq^N)+\sum_{s=1}^{N-1}(-1)^{s}
{\bf T}_{\Lambda_1+\cdots+
\Lambda_s}(tq^{-1}){\bf Q}_i(tq^{N-2s})+(-1)^{N}{\bf Q}_i(tq^{-N})=0,
\label{rel:BaxTQ1}\\
&&
\overline{\bf Q}_i(tq^{-N})+\sum_{s=1}^{N-1}(-1)^{s}
\overline{\bf T}_{\Lambda_1+\cdots+\Lambda_s}(tq)
\overline{\bf Q}_i(tq^{-N+2s})+(-1)^{N}
\overline{\bf Q}_i(tq^{N})=0,
\label{rel:BaxTQ2}
\end{eqnarray}
with $i=1,2,\cdots,N$.
This Baxter's $T$-$Q$ relation, (\ref{rel:BaxTQ1}) and 
(\ref{rel:BaxTQ2}), coincides with those in \cite{BHK} 
upon $N=3$ specialization.
Note that the specialization to $N=2$
does not yield the formulae in
\cite{BLZ1, BLZ2, BLZ3},
because the Dynkin-diagram for $N=2$
is different from those for $N \geqq 3$.
We have to give separate definitions of the bosons, 
the $q$-oscillator
and the screening currents for $N=2$,
\cite{BLZ1, BLZ2, BLZ3}.
This Baxter's $T$-$Q$ relation
(\ref{rel:BaxTQ1}), (\ref{rel:BaxTQ2})
coincides with those of \cite{KLWZ} for $N \geqq 3$.
In \cite{KLWZ},
I.Krichever, O.Lipan, P.Wiegmann and A.Zabrodin
gave the conjecture that
the standard
objects of quantum integrable models are identified
with elements of classical nonlinear integrable
difference equation.
For simplest example they 
showed that the fusion rules for quantum transfer 
matrices coincide with the Hirota-Miwa's
bilinear difference equation \cite{H,M}
(the discrete KP).
They derived
higher-rank generalization of Baxter's
$T$-$Q$ relation by analysing
the Hirota-Miwa's bilinear difference equation
(classical nonlinear integrable
difference equation), too.
In this paper,
we derive the same Baxter's $T$-$Q$ relation
by analysing the quantum field theory of the KP
(quantum integrable model).
Hence this paper
give a supporting argument
of the conjecture 
on quantum and classical-discrete integrable models,
by 
I.Krichever, O.Lipan, P.Wiegmann and A.Zabrodin
\cite{KLWZ}.
As the consequence of (\ref{rel:det3}) and (\ref{rel:det4}),
we have the bilinear formulae of the $T$-operator 
(\ref{rel:qua1}) and (\ref{rel:qua2}).
\begin{eqnarray}
(-1)^{\frac{(N-1)(N-2)}{2}}c_0
{\bf T}_{m\Lambda_1}(t)&=&\sum_{s=1}^N 
(-1)^{s+1}c_s 
{\bf Q}_s(tq^{2m+N-1})\overline{\bf Q}_s(tq^{-1}),
\label{rel:qua1}\\
(-1)^{\frac{(N-1)(N-2)}{2}}c_0
\overline{\bf T}_{m\Lambda_1}(t)&=&
\sum_{s=1}^N (-1)^{s+1}c_s 
\overline{\bf Q}_s(tq^{-2m-N+1}){\bf Q}_s(tq),
\label{rel:qua2}
\end{eqnarray}
and
\begin{eqnarray}
(-1)^{\frac{(N-1)(N-2)}{2}}c_0
{\bf T}_{m(\Lambda_1+\cdots+\Lambda_{N-1})}(t)&=&
\sum_{s=1}^N (-1)^{N+s}c_s 
\overline{\bf Q}_s(tq^{-2m-1})
\overline{\bf Q}_s(tq^{N-1}),
\label{rel:qua3}\\
(-1)^{\frac{(N-1)(N-2)}{2}}c_0
\overline{\bf T}_{m(\Lambda_1+\cdots+\Lambda_{N-1})}
(t)&=&\sum_{s=1}^N (-1)^{N+s}
c_s {\bf Q}_s(tq^{2m+1})
\overline{\bf Q}_s(tq^{-N+1}).
\label{rel:qua4}
\end{eqnarray}
As a consequence of the determinant formulae
(\ref{rel:det1}) and (\ref{rel:det2}),
we have the Jacobi-Trudi formulae of the $T$-operator.
For the Young-diagram 
$\mu=(\mu_1,\mu_1,\cdots,\mu_{N-1},0)$, we have
\begin{eqnarray}
{\bf T}_\mu(t)=\left|\begin{array}{ccccc}
{\bf \tau}^{(\mu_1')}(t)&\cdots&
{\bf \tau}^{(\mu_1'+j-1)}(tq^{2(j-1)})&\cdots&
{\bf \tau}^{(\mu_1'+l(\mu')-1)}(tq^{2(l(\mu')-1)})\\
\cdots&\cdots&\cdots&\cdots&\cdots\\
{\bf \tau}^{(\mu_i'-i+1)}(t)&\cdots&
{\bf \tau}^{(\mu_i'-i+j)}(tq^{2(j-1)})&\cdots&
{\bf \tau}^{(\mu_i'-i+l(\mu'))}(tq^{2(l(\mu')-1)})\\
\cdots&\cdots&\cdots&\cdots&\cdots\\
{\bf \tau}^{(\mu_{l(\mu')}'-l(\mu')+1)}(t)&\cdots&
{\bf \tau}^{(\mu_{l(\mu')}'-l(\mu')+j)}(tq^{2(j-1)})&\cdots&
{\bf \tau}^{(\mu_{l(\mu')}')}(tq^{2(l(\mu')-1)})
\end{array}\right|.\nonumber\\
\label{rel:JT}
\end{eqnarray}
Here we have set 
$\mu'=(\mu_1',\mu_2',\cdots,\mu_N')$ the transpose
Young-diagram of $\mu$, and $l(\mu')=\mu_1$.
We have set
$\tau^{(s)}(t)={\bf T}_{\Lambda_1+\cdots+\Lambda_s}(t)$.
We have $\tau^{(0)}(t)=\tau^{(N)}(t)=1$.
The above conjecturous functional relations
of the $T$-$Q$ operators,
(\ref{rel:det1}), 
(\ref{rel:det2}), 
(\ref{rel:det1'}), 
(\ref{rel:det2'}), 
(\ref{rel:det3}), 
(\ref{rel:det4}), 
(\ref{rel:BaxTQ1}),
(\ref{rel:BaxTQ2}),
(\ref{rel:qua1}), 
(\ref{rel:qua2}), 
(\ref{rel:qua3}), 
(\ref{rel:qua4}), 
(\ref{rel:JT}),
coincide with the previous work
\cite{BHK} upon $N=3$ specialization.

\subsection{Conclusion}

In this paper we present $q$-oscillator representation
of the Borel subalgebra $U_q'(\widehat{sl_N})$, (\ref{def:q-osc}).
By using this $q$-oscillator representation,
we give the free field realization of the Baxter's $Q$-operator
${\bf Q}_j(t), \overline{\bf Q}_j(t)$ with $j=1,2,\cdots,N$, for 
the $W_N$-algebra,
(\ref{def:A1}), (\ref{def:A2}), (\ref{def:Q}).
The commutativity of the $Q$-operator is direct 
consequence of the Yang-Baxter equation.
We give conjecturous determinant formulae of
the $T$-$Q$ operator for the $W_N$-algebra, (\ref{rel:det1}), 
(\ref{rel:det2}), (\ref{rel:det3}), 
(\ref{rel:det4}), 
which produce the higher-rank $W_N$-generalization of 
the Baxter's $T$-$Q$ relation, 
(\ref{rel:BaxTQ1}), (\ref{rel:BaxTQ2}).
We have checked these determinant formulae of
the $T$-$Q$ operator, (\ref{rel:det1}), 
(\ref{rel:det2}), (\ref{rel:det3}), 
(\ref{rel:det4}) up to the order $O(t^2)$ in appendix.
Because the scheme of funtional relations works well,
we conclude that the number of the $Q$-operators for 
the $W_N$-algebra, is just $2N$, $(N \geqq 3)$.
In this paper we didn't give complete proof of the
determinant formulae for the $W_N$-algebra.  
V.Bazhanov, A.Hibberd and S.Khoroshkin \cite{BHK}
gave proof of the determinant
formulae for the $W_3$-algebra.
Their proof is based on
the trace of the universal ${\cal L}$-operator
over Verma module, and the Bernstein-Gel'fand-Gel'fand
(BGG) resolution.
Because we have already established 
conjecturous determinant formulae,
higher-rank generalization of complete proof seems
calculation problem. However it is not so easy.

~\\\\
{\bf Acknowledgements}\\
The author would like to thank Prof.V.Bazhanov and Prof.M.Jimbo
for useful communications.
The author would like to thank Institute of Advanced Studies,
Australian National University
for the hospitality during his visit to
Canberra in March 2008.
The author would like to thank 
Prof.P.Bouwknegt,
Prof.A.Chervov, 
Prof.V.Gerdjikov,
Prof.K.Hasegawa and
Prof.V.Mangazeev  
for their interests in this work.
This work is partly supported by
Grant-in Aid for Young Scientist
{\bf B} (18740092) from JSPS.

\begin{appendix}

\section{Supporting Arguments}

In this appendix we give some supporting arguments
on conjecturous formulae of the determinant formulae
(\ref{rel:det1}), (\ref{rel:det2}), 
(\ref{rel:det1'}), (\ref{rel:det2'}), 
(\ref{rel:det3}), (\ref{rel:det4}).
We check those determinant formulae up to the order $O(t^2)$.
At first we prepare the Taylor expansion
of ${\bf A}_j(t), \overline{\bf A}_j(t)$.
Let us set $\pi_j=\overline{\pi}_j=
\pi_1^+\otimes \cdots \otimes \pi_{N-1}^+$.
Taking the trace for the basis $\{|n_1,n_2,\cdots, n_{N-1}\rangle
=({\cal H}_1^*)^{n_1}({\cal H}_2^*)^{n_2}\cdots
({\cal H}_{N-1}^*)^{n_{N-1}}|0\rangle_+ \otimes \cdots \otimes |0\rangle_+\}_{n_1,n_2,
\cdots, n_{N-1} \in {\mathbb N}}$, we have
\begin{eqnarray} 
Z_j(t)={\rm Tr}_{\pi_j}\left(
\exp\left(-2\pi i \sqrt{\frac{r^*}{r}}
\sum_{k=1}^{N-1}(P_{\bar{\epsilon}_{j}}-P_{\bar{\epsilon}_{j+k}})
\otimes {\cal H}_k\right)
\right)=
\prod_{k=1
\atop{k \neq j}}^N
\left(1-\frac{z_k}{z_{j}}\right)^{-1},\nonumber
\end{eqnarray}
with $j=1,2,\cdots, N$.
As the same manner as the above,
we have
\begin{eqnarray}
\overline{Z}_j(t)
=
\prod_{k=1
\atop{k \neq j}}^N\left(1-\frac{z_{j}}{z_{k}}\right)^{-1}.
\nonumber
\end{eqnarray}
Let us set $a_i, \overline{a}_i$ by
\begin{eqnarray}
{\bf A}_i(t)=1+a_it+O(t^2),~~
\overline{\bf A}_i(t)=1+\overline{a}_i t+O(t^2).\nonumber
\end{eqnarray}
Let us set
\begin{eqnarray}
{\it J}^{(n)}_{k_1,k_2,\cdots,k_N}
&=&\int \cdots \int_{2\pi \geqq u_1 \geqq u_2 \geqq \cdots \geqq u_{N n}
\geqq 0}
V_{k_1}(u_1)V_{k_2}(u_2)\cdots V_{k_N}(u_N)\nonumber\\
&\times&
V_{k_1}(u_{N+1})V_{k_2}(u_{N+2})\cdots V_{k_N}(u_{2N})\cdots
\nonumber\\
&\times&
V_{k_1}(u_{N(n-1)+1})V_{k_2}(u_{N(n-1)+2})\cdots V_{k_{Nn}}(u_{N n})
du_1 du_2 \cdots du_{N n}.\nonumber
\end{eqnarray}
Let us calculate coefficient of ${\cal J}_{1,2,\cdots,N}^{(1)}$ in $a_i$.
We have
\begin{eqnarray}
&&
{\rm Tr}_{\pi_i}\left(
o_{t,i}\left(
\exp\left(-2\pi i \sqrt{\frac{r^*}{r}}\sum_{k=1}^{N-1}P_{\omega_k}\otimes h_k \right)
e_{1} e_2 e_{3} \cdots e_{N}\right)\right)\times
{\cal J}_{1,2,\cdots,N}^{(1)}\nonumber\\
&=&t(q-q^{-1})^{N-2}q^{\frac{N-2}{2}}
\nonumber\\
&\times&\prod_{k=1}^{N-1}
{\rm Tr}_{\pi^+_k}
\left(\exp\left(-2\pi i \sqrt{\frac{r^*}{r}}
(P_{\bar{\epsilon}_{i}}-P_{\bar{\epsilon}_{i+k}})
\otimes {\cal H}_k\right) {\cal E}_k {\cal E}_k^*\right)
\times
{\cal J}_{1,2,\cdots,N}^{(1)}.\nonumber
\end{eqnarray}
Taking the trace and
dividing $Z_i(t)$, we have
\begin{eqnarray}
a_{i}&=&
\frac{
q^{\frac{3}{2}N-2}z_i^{N-2}z_1
}{\displaystyle (q-q^{-1}) 
\prod_{k=1
\atop{k \neq i}}^N
(q^2z_i-z_k)}\times
{\cal J}_{1,2,\cdots,N}^{(1)}+\cdots,\nonumber
\end{eqnarray}
with $i=1,2,\cdots,N$.
As the same manner as the above, we have
\begin{eqnarray}
\overline{a}_{i}&=&
(-1)^N
\frac{
q^{\frac{1}{2}N}z_i^{N-2}z_1
}{\displaystyle (q-q^{-1}) 
\prod_{k=1
\atop{k \neq i}}^N
(q^2z_k-z_i)}\times
{\cal J}_{1,2,\cdots,N}^{(1)}+\cdots,\nonumber
\end{eqnarray}
with $i=1,2,\cdots,N$.
Let us check the determinant relations between
${\bf Q}_i(t)$ and $\overline{\bf Q}_i(t)$,
(\ref{rel:det3}) and (\ref{rel:det4}).
We have
\begin{eqnarray}
&&\left|
\begin{array}{cccc}
{\bf Q}_1(tq^{N-2})&{\bf Q}_1(tq^{N-4})&\cdots&
{\bf Q}_1(tq^{-N+2})\\
\cdots&\cdots&\cdots&\cdots\\
{\bf Q}_{i-1}(tq^{N-2})&
{\bf Q}_{i-1}(tq^{N-3})&\cdots&
{\bf Q}_{i-1}(tq^{-N+2})\\
{\bf Q}_{i+1}(tq^{N-2})&
{\bf Q}_{i+1}(tq^{N-3})&\cdots&
{\bf Q}_{i+1}(tq^{-N+2})\\
\cdots&\cdots&\cdots&\cdots
\\
{\bf Q}_N(tq^{N-2})&{\bf Q}_N(tq^{N-4})&\cdots&
{\bf Q}_N(tq^{-N+2})
\end{array}
\right|\nonumber\\
&=&
t^{-\sqrt{\frac{r}{r^*}}P_{\bar{\epsilon}_i}}
\prod_{1\leqq j<k \leqq N
\atop{j,k\neq i}}
\left(\sqrt{\frac{z_j}{z_k}}-
\sqrt{\frac{z_k}{z_j}}\right)
\left(
1+\left(
\sum_{j=1
\atop{j \neq i}}^N
a_j 
\prod_{k=1\atop{k \neq i,j}}^N
\frac{(q^2z_j-z_k)}
{(z_j-z_k)}q^{-N+2}\right)t+O(t^2)\right).\nonumber
\end{eqnarray}
Inserting the formulae of $a_i$ into RHS and using the following identity,
\begin{eqnarray}
\sum_{j=1
\atop{j \neq i}}^N
\frac{z_j^{N-2}}{
\displaystyle
(z_jq^2-z_i)
\prod_{k=1\atop{k\neq i,j}}^N
(z_j-z_k)}=(-1)^N \frac{z_i^{N-2}}{
\displaystyle
\prod_{k=1
\atop{k \neq i}}^N
(z_kq^2-z_i)}.\nonumber
\end{eqnarray}
we have
\begin{eqnarray}
t^{-\sqrt{\frac{r}{r^*}}P_{\bar{\epsilon}_i}}
c_i
\left(
1+(-1)^N \frac{q^{\frac{1}{2}N}z_i^{N-2}z_1}{
\displaystyle
(q-q^{-1})
\prod_{k=1\atop{k \neq i}}^N
(z_kq^2-z_i)
}\times
{\cal J}_{1,2,\cdots,N}^{(1)}\times t+\cdots
\right),\nonumber
\end{eqnarray}
which coincides with leading terms of
$\overline{\bf Q}_i(t)$.
As the same argument as the above,
the coefficients of
${\cal J}_{k_1,k_2,\cdots,k_N}^{(1)}$ 
coincide with each other up to the order $O(t^2)$.
Now 
we have checked 
the determinant formulae (\ref{rel:det3}) and 
(\ref{rel:det4}) up to the order $O(t^2)$.
For the second we check the quantum Wronskian condition
(\ref{rel:det1'}) and 
(\ref{rel:det2'}) up to the order $O(t^2)$.
We have Taylor expansion of
determinant of ${\bf Q}_j(t)$,
\begin{eqnarray}
&&\left|
\begin{array}{cccc}
{\bf Q}_1(tq^{N-1})&{\bf Q}_1(tq^{N-3})&\cdots&{\bf Q}_1(tq^{-N+1})\\
{\bf Q}_2(tq^{N-1})&{\bf Q}_2(tq^{N-3})&\cdots&{\bf Q}_2(tq^{-N+1})\\
\cdots&\cdots&\cdots&\cdots
\\
{\bf Q}_N(tq^{N-1})&{\bf Q}_N(tq^{N-3})&\cdots&{\bf Q}_N(tq^{-N+1})
\end{array}
\right|\nonumber\\
&=&
\prod_{1\leqq j<k \leqq N}
\left(\sqrt{\frac{z_j}{z_k}}-
\sqrt{\frac{z_k}{z_j}}\right)
\left(
1+\left(
\sum_{i=1}^N
a_i \prod_{k=1
\atop{k \neq i}}^N
\frac{(q^2z_i-z_k)}{(z_i-z_k)}q^{-N+1}
\right)t+O(t^2)\right).\nonumber
\end{eqnarray}
Insrting the explicit formulae of $a_i$ into RHS 
and using the following identity,
\begin{eqnarray}
\sum_{i=1}^N
(-1)^{i+1}z_i^{N-2}
\prod_{1\leqq j<k \leqq N
\atop{j,k\neq i}}(z_j-z_k)=0,\nonumber
\end{eqnarray}
we have
\begin{eqnarray}
\prod_{1\leqq j<k \leqq N}
\left(\sqrt{\frac{z_j}{z_k}}-
\sqrt{\frac{z_k}{z_j}}\right)
\left(
1+O(t^2)\right).\nonumber
\end{eqnarray}
Now we have checked the quantum Wronskian condition 
(\ref{rel:det1'})
and (\ref{rel:det2'})
up to the order 
$O(t^2)$.
Next we consider the determinant formulae
(\ref{rel:det1}) and (\ref{rel:det2})
for the special cases 
$\mu=\Lambda_1$ and $\mu=\Lambda_1+\cdots+\Lambda_{N-1}$.
Because we have checked the formulae
(\ref{rel:det3}) and (\ref{rel:det4}) 
up to the order $O(t^2)$,
it is enough to show (\ref{rel:qua1}), 
(\ref{rel:qua2}), (\ref{rel:qua3}) and (\ref{rel:qua4}) 
in order to
show (\ref{rel:det1}) and (\ref{rel:det2}) 
up to the order $O(t^2)$.
We have
\begin{eqnarray}
&&(-1)^{\frac{(N-1)(N-2)}{2}}
\sum_{s=1}^N (-1)^{s+1}c_s 
{\bf Q}_s(tq^{N+1})\overline{\bf Q}_s(tq^{-1})\nonumber\\
&=&
\sum_{s=1}^N
(-1)^{s+1}
z_s^N 
\prod_{1\leqq j<k\leqq N
\atop{j,k \neq s}}(z_j-z_k)
\left(1+t(
q^{N+1}a_s+q^{-1}\overline{a}_s)+O(t^2)\right).\nonumber
\end{eqnarray}
Using the following relation,
\begin{eqnarray}
\sum_{s=1}^N
(-1)^{s+1}
z_s^N 
\prod_{1\leqq j<k\leqq N
\atop{j,k \neq s}}(z_j-z_k)=
\prod_{1\leqq j<k\leqq N}(z_j-z_k)
(z_1+z_2+\cdots+z_N),\nonumber
\end{eqnarray}
we show that the first leading term becomes 
$c_0 \sum_{j=1}^Nz_j.$
Inserting the explicit formulae of 
$a_s,
\overline{a}_s$ and using the following relation,
\begin{eqnarray}
\sum_{s=1}^N\frac{z_s^{2N-2}}{
\displaystyle
\prod_{k=1\atop{k\neq s}}^N (z_s-z_k)}
\left(\frac{q}{
\displaystyle
\prod_{k=1\atop{k \neq s}}^N(z_s-q^{2}z_k)}-
\frac{q^{-1}}{
\displaystyle
\prod_{k=1\atop{k \neq s}}^N(z_s-q^{-2}z_k)}
\right)=(q-q^{-1})\nonumber
\end{eqnarray}
we have
the second leading term,
\begin{eqnarray}
t q^{\frac{Nn}{2}} \prod_{1\leqq j<k \leqq N}(z_k-z_j)
\left(
z_1 {\cal J}_{1,2,\cdots,N}^{
(1)}+z_2 {\cal J}_{2,3,\cdots,N,1}^{(1)}+\cdots \right).\nonumber
\end{eqnarray}
Now we need explicit formulae of
${\bf T}_{\Lambda_1}(t)$.
Let us fix a basis 
of the irreducible highest representation of $U_q(gl_N)$
with $\Lambda_1$ by
\begin{eqnarray}
|\Lambda_1\rangle, 
\pi^{(\Lambda_1)}(E_{\alpha_1})|\Lambda_1\rangle,
\pi^{(\Lambda_1)}(E_{\alpha_2}E_{\alpha_1})|\Lambda_1\rangle,
\cdots,
\pi^{(\Lambda_1)}(E_{\alpha_{N-1}}\cdots
E_{\alpha_1})|\Lambda_1\rangle,\nonumber
\end{eqnarray}
The matrix representation of $\pi^{(\Lambda_1)}$
are written upon this basis by
\begin{eqnarray}
&&\pi^{(\Lambda_1)}(E_{\alpha_i})=(
\delta_{j,i}
\delta_{k,i+1})_{1\leqq j,k \leqq N},~~(1\leqq i \leqq N-1),\nonumber\\
&&
\pi^{(\Lambda_1)}(F_{\alpha_i})=(\delta_{j,i+1}
\delta_{k,i})_{1\leqq j,k \leqq N},~~(1\leqq i \leqq N-1),
\nonumber\\
&&\pi^{(\Lambda_1)}(H_{i})=
(\delta_{j,i}\delta_{k,i})
_{1\leqq j,k \leqq N},~(1\leqq i \leqq N).\nonumber
\end{eqnarray}
Using this matrix representation, we have
\begin{eqnarray}
{\bf T}_{\Lambda_1}(t)&=&
\sum_{j=1}^N z_j
+\sum_{n=1}^\infty
t^n q^{\frac{N n}{2}}\sum_{j=1}^N z_{j} 
{\it J}_{j,j+1,j+2,\cdots,j+N-1}^{(n)}.\nonumber
\end{eqnarray}
Now we have checked the determinant formulae
(\ref{rel:det1}) for $\mu=\Lambda_1$ up to the order
$O(t^2)$.
As the same manner we checked
 the determinant formulae
(\ref{rel:det1}) for 
$\mu=\Lambda_1+\cdots+\Lambda_{N-1}$ and
(\ref{rel:det2}) for 
$\mu=\Lambda_1,
\Lambda_1+\cdots+\Lambda_{N-1}$,
up to the order
$O(t^2)$.
For reader's convenience we sumarize the explicit formulae
of
${\bf T}_{\Lambda_1+\cdots+\Lambda_{N-1}}(t)$,
$\overline{\bf T}_{\Lambda_1}(t)$
and
$\overline{\bf T}_{\Lambda_1+\cdots+\Lambda_{N-1}}(t)$.
The matrix representation of $\pi^{(\Lambda_1+\cdots+\Lambda_{N-1})}$
are written by
\begin{eqnarray}
&&\pi^{(\Lambda_1+\cdots+\Lambda_{N-1})}(E_{\alpha_i})=(
\delta_{j,N-i}
\delta_{k,N-i+1})_{1\leqq j,k \leqq N},~~(1\leqq i \leqq N-1),\nonumber\\
&&
\pi^{(\Lambda_1+\cdots+\Lambda_{N-1})}(F_{\alpha_i})=(\delta_{j,N-i+1}
\delta_{k,N-i})_{1\leqq j,k \leqq N},~~(1\leqq i \leqq N-1),
\nonumber\\
&&\pi^{(\Lambda_1+\cdots+\Lambda_{N-1})}(H_{i}-H_{i+1})=
(\delta_{j,N-i}\delta_{k,N-i}-\delta_{j,N-i+1}\delta_{k,N-i+1})
_{1\leqq j,k \leqq N},~(1\leqq i \leqq N-1).\nonumber
\end{eqnarray}
We have
\begin{eqnarray}
{\bf T}_{\Lambda_1+\cdots+\Lambda_{N-1}}(t)&=&
\sum_{j=1}^N \frac{1}{z_j}
+\sum_{n=1}^\infty
t^n q^{(2-\frac{N}{2})n}
(-1)^{Nn}\sum_{j=1}^N \frac{1}{z_{j}} 
{\it J}_{j+N-1,\cdots ,j+1,j}^{(n)},\nonumber\\
\overline{\bf T}_{\Lambda_1}(t)&=&
\sum_{j=1}^N z_{j} 
+\sum_{n=0}^\infty
t^n q^{-\frac{nN}{2}}\sum_{j=1}^N z_{j} 
{\it J}_{j,j+1,j+2,\cdots,j+N-1}^{(n)},\nonumber\\
\overline{\bf T}_{\Lambda_1+\cdots+\Lambda_{N-1}}(t)&=&
\sum_{j=1}^N \frac{1}{z_j}
+\sum_{n=1}^\infty t^n q^{(\frac{N}{2}-2)n}
\sum_{j=1}^N \frac{1}{z_{j}} 
{\it J}_{j+N-1,\cdots ,j+1,j}^{(n)}.\nonumber
\end{eqnarray}
Using these explicit formulae, we have
\begin{eqnarray}
{\bf T}_{\Lambda_1}(q^{-\frac{N}{2}}t)&=&
\overline{\bf T}_{\Lambda_{1}}(q^{\frac{N}{2}}t),\nonumber\\
{\bf T}_{\Lambda_1+\cdots+\Lambda_{N-1}}(q^{\frac{N-4}{2}}t)&=&
\overline{\bf T}_{\Lambda_1+\cdots+\Lambda_{N-1}}(q^{\frac{4-N}{2}}t),\nonumber\\
{\bf T}_0(t)&=&\overline{\bf T}_0(t)=1.\nonumber
\nonumber
\end{eqnarray}
\end{appendix}

\end{document}